\documentclass[epj]{webofc}
\usepackage[varg]{txfonts}   
\usepackage{graphicx,color} 
\usepackage{bm}       
\usepackage{amsmath}  
\usepackage{amssymb}
\woctitle{21st International Conference on Few-Body Problems in Physics}
\def\s#1{{\scriptscriptstyle #1}}
\def\spr{\!\cdot\!}

\def\noeq#1{(\ref{#1})}
\def\1eq#1{Eq.~(\ref{#1})}

\def\2eqs#1#2{Eqs.~(\ref{#1}) and~(\ref{#2})}
\def\3eqs#1#2#3{Eqs.~(\ref{#1}),~(\ref{#2}) and~(\ref{#3})}

\begin{document}
\title{From continuum QCD to hadron observables}
\author{Daniele Binosi\inst{1}\fnsep\thanks{\email{binosi@ectstar.eu}}}
       
\institute{European Centre for Theoretical Studies in Nuclear Physics and Related Areas (ECT*) \\ and Fondazione Bruno Kessler, Villa Tambosi, Strada delle Tabarelle 286, I-38123 Villazzano (TN)  Italy}

\abstract{
  We show that the form of the renormalization group invariant quark-gluon interaction predicted by a refined nonperturbative analysis of the QCD gauge sector is in quantitative agreement with the one required for describing a wide range of hadron observables using sophisticated truncation schemes of the Schwinger-Dyson equations relevant in the matter sector.    
}
\maketitle

\section{\label{sec-1}Introduction}

The quark gap equation is arguably one of the most important equations appearing in QCD. Solutions to this equation must describe the evolution of a (chiral) current quark of perturbative QCD into a constituent quark with a mass around 350 MeV. This nonperturbative effect thusly generates mass from nothing: being responsible for approximately 98\% of the proton's mass, it represents the most important mass generating mechanism for visible matter in the Universe (the Higgs is almost irrelevant for light quarks)~\cite{Roberts:2015dea}.  

Writing for the quark propagator $S(p)=1/[iA(p^2)\gamma\spr p+B(p^2)]$, the gap equation reads
\begin{align}
	S^{-1}(p)=Z_2(i\gamma\spr p+m_0)+\frac43Z_1\int_qg^2\Delta_{\mu\nu}(k)\gamma^\mu S(q)\Gamma^\nu(q,p),
\end{align}
where $k=p-q$, $m_0$ is the quark bare mass, $Z_{1,2}$ the vertex and wave-function renormalization constants, $\Delta_{\mu\nu}=(g_{\mu\nu}-k_\mu k_\nu/k^2)\Delta(k^2)$ the (Landau gauge) gluon propagator and $\Gamma^\nu$ the quark gluon vertex. Owing to asymptotic freedom, the model input for realistic studies of this equation can be reduced to providing a statement on the nature of the gap equation kernel in the momentum region $k^2\lesssim2$ GeV$^2$, as above this region perturbation theory takes over. In practice one writes
\begin{align}
	Z_1g^2\Delta_{\mu\nu}(k)\Gamma^\nu(q,p)={\cal I}(k^2)\Delta^\s{\mathrm{free}}_{\mu\nu}(k)Z_2\Gamma^\nu_\s{\mathrm{A}}(q,p).
\end{align} 
where ${\cal I}$ represents the so-called quark-gluon interaction strength, $\Delta^\s{\mathrm{free}}_{\mu\nu}=(g_{\mu\nu}-k_\mu k_\nu/k^2)/k^2$ is the tree-level gluon propagator and, finally, $\Gamma^\nu_\s{\mathrm{A}}$ represents a suitable vertex Ansatz reducing to $Z_2\gamma^\nu$ in the perturbative region $k^2>2$ GeV$^2$.

There are now two possible approaches that can be employed for determining the interaction strength ${\cal I}$. In a {\it bottom-up} approach, this quantity is obtained by fitting bound state properties data within a well-defined truncation scheme of the relevant equations (see, e.g.,~\cite{Eichmann:2008ef}). In a {\it top-down} approach, one attempts instead to directly compute ${\cal I}$ by means of {\it ab-initio} studies of the QCD gauge and ghost sectors.

The widespread phenomenological success of the former approach, together with the significant progress made over the last decade in our understanding of the QCD infrared physics from first principles-- in particular the consensus that the gauge sector is characterized by the dynamical generation of a gluon mass-scale~\cite{Cornwall:1981zr,Aguilar:2008xm}--, raises an important question~\cite{Binosi:2014aea}: Is the interaction strength evaluated within these two approaches in qualitative and (possibly) quantitative agreement?

\section{\label{sec-2}Bottom-up approach}

As reviewed elsewhere~\cite{Chang:2011vu,Bashir:2012fs,Cloet:2013jya} successful explanations and predictions of numerous hadrons observables can be obtained choosing  
\begin{equation}
{\cal I}(k^2)=k^2 {\cal G}(k^2);\quad
{\cal G}(k^2)=\frac{8\pi^2}{\omega^4}D{\mathrm e}^{-k^2/\omega^2}+\frac{8\pi^2\gamma_m (1-{\mathrm e}^{-k^2/4m_t^2})}{k^2\ln[\tau+(1+k^2/\Lambda_\s{\rm QCD}^2)^2]},
\label{bu}
\end{equation}
where $\gamma_m=12/(33-2N_f)$ [typically, $N_f=4$], $\Lambda_\s{\rm QCD}=0.57$ GeV (in the momentum subtraction scheme); \mbox{$\tau={\rm e}^2-1$}, $m_t=0.5$ GeV. 

At a first glance the bottom-up interaction strength~\noeq{bu} depends on two parameters: $D$ and $\omega$. However, explicit computations have shown that as long as the two parameters are related through
$D\omega=(\varsigma_\s{\mathrm{G}})^3=\mathrm{const}$ and $\omega\in[0.4,0.6]$ GeV, one can reproduce a large body of observable properties of ground-state vector- and isospin-nonzero pesudoscalar mesons as well as numerous properties of the nucleon and $\Delta$ resonance~\cite{Qin:2011dd}. The value of $\varsigma_\s{\mathrm{G}}$ is then fixed by the requirement of reproducing the correct value of the pion decay constant $f_\pi$, with the resulting value depending on the form of the vertex Ansatz employed. 

In the case of the rainbow-ladder truncation, corresponding to $\Gamma^\nu_\s{\mathrm{A}}\sim\gamma_\nu$, one obtains~$\varsigma_\s{\mathrm{RL}}=0.87$ GeV, whereas the improved truncation scheme of~\cite{Chang:2009zb} (which incorporates dynamical chiral symmetry breaking nonperturbatively in the bound-states integral equation through an accurate representation of the dressed quark-gluon vertex) gives $\varsigma_\s{\rm DB}=0.55$ GeV.

Thus, the question we originally posed can be rephrased as follows: does a top-down approach produce a shape for the interaction strength qualitatively consistent with the one found in a bottom up approach? and if so, is there a value of~$\varsigma_\s{\mathrm{G}}$ turning the agreement quantitative?

\section{\label{sec-3}Top-down approach}

The top-down interaction strength can be obtained within the so-called pinch-technique background field method framework~\cite{Aguilar:2006gr,Binosi:2007pi,Binosi:2008qk,Binosi:2009qm} and is found to be~\cite{Binosi:2014aea}
\begin{equation}
{\cal I}(k^2)=k^2 \widehat{d}(k^2);\qquad \widehat{d}(k^2)=\frac{\alpha_s(\mu^2)\Delta(k^2;\mu^2)}{[1+G(k^2;\mu^2)]^2},
\label{td}
\end{equation}
with $\alpha_s$ the strong coupling evaluated at the renormalization point $\mu^2$, and~$G$ is one of two possible form factors (the other one being $L$) of a certain function $\Lambda_{\mu\nu}=g_{\mu\nu}G+(k_\mu k_\nu/k^2)L$ describing the ghost-gluon dynamics and typically appearing in the PT-BFM framework. BRST symmetry implies then that, in the Landau gauge, these two form factors are related to the ghost dressing function $F$ (defined as $k^2$ times the ghost propagator) through $F^{-1}(k^2)=1+G(k^2)+L(k^2)$~\cite{Binosi:2013cea}. It is worth noticing that $\widehat{d}$ is a renormalization group invariant quantity~\cite{Aguilar:2009nf}, and does not depend on the valence-quark content of the corresponding Bethe-Salpeter equation; in addition~\1eq{td} is parameter free.

The procedure of constructing the top-down interaction involves the following main steps~\cite{Aguilar:2009nf,Binosi:2014aea}. One begins with the lattice QCD results for the gluon propagator renormalized at a suitable value $\mu^2$, and use them as input in the gap equation satisfied by the ghost dressing function $F$. The latter equation is then solved and the strong coupling $\alpha_s$ determined so that the resulting solution best matches the available lattice data. From the obtained $F$ and $\alpha_s$, one can next determine the function $G$ (or, equivalently, $L$), and therefore construct the final quantity $\widehat{d}$. The renormalization group invariance can be checked by changing the initial renormalization point $\mu^2$ of the gluon propagator, and observing that ({\it i}) the resulting curves for $\widehat{d}$ precisely overlap, and ({\it ii}) that this happens for the correct values of the strong coupling~$\alpha_s(\mu^2)$~\cite{Binosi:2014aea}.  

The major source of uncertainty in this procedure is related to the determination of $G$ (or $L$) from the ghost dressing function $F$, as this entails a modelling of the ghost-gluon vertex. A (gross) estimate of this uncertainty can be obtained by first evaluating the form factor $L$ from its Schwinger-Dyson equation, and then use the identity between $F,G$ and $L$ with the replacement $L\to\delta L$ and $\delta\in[0,1]$. $\delta=1$ corresponds to the $1+G$ obtained through the solution of the corresponding Schwinger-Dyson equation, while $\delta=0$ corresponds to the often employed approximation $1+G\approx1/F$ ($L$ is known to be a subleading term with $L(0)=0$). 

\section{\label{sec-4}Comparison}

\begin{figure}[t]
\centerline{\includegraphics[width=0.477\linewidth]{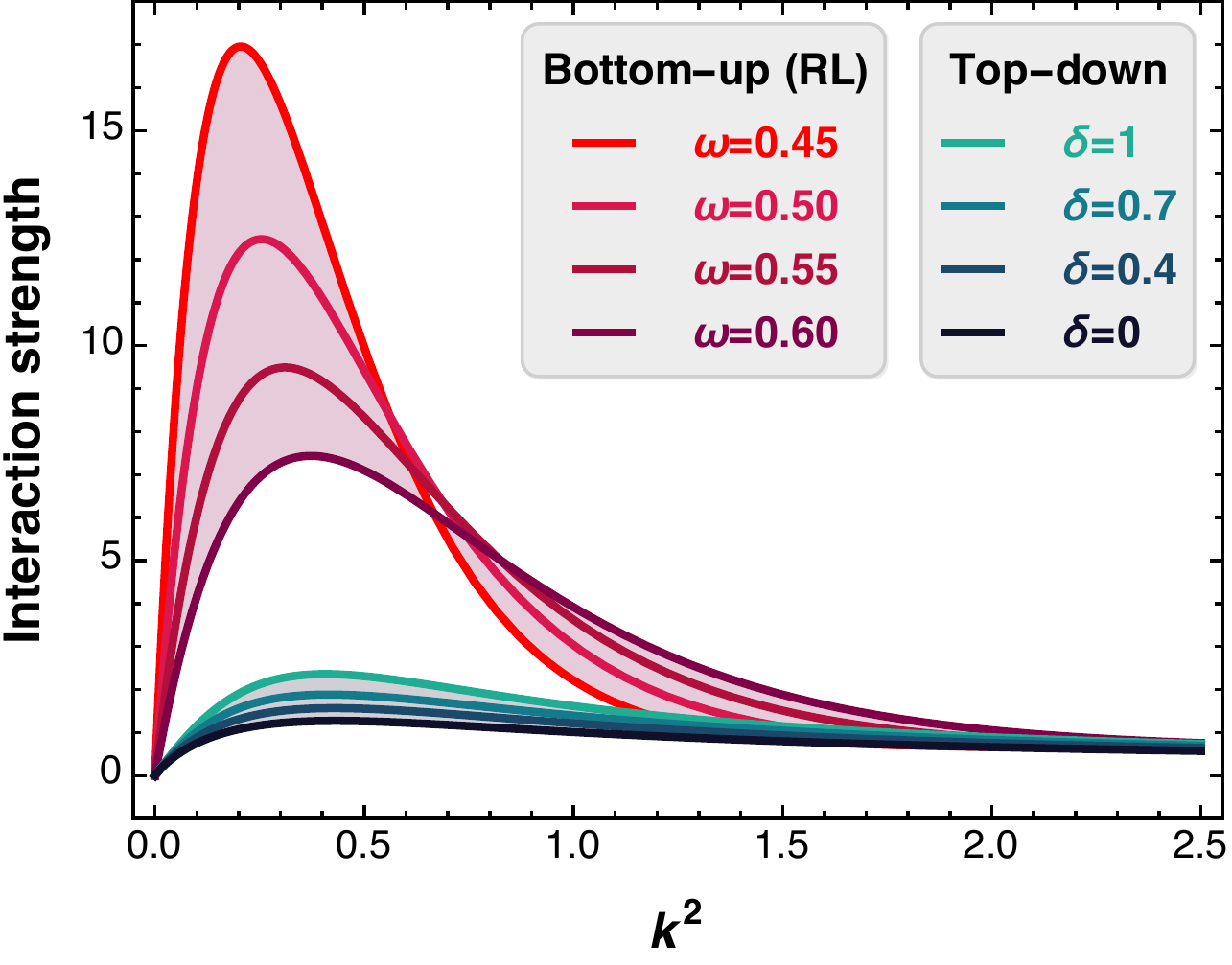}\hspace{0.4cm}
\includegraphics[width=0.469\linewidth]{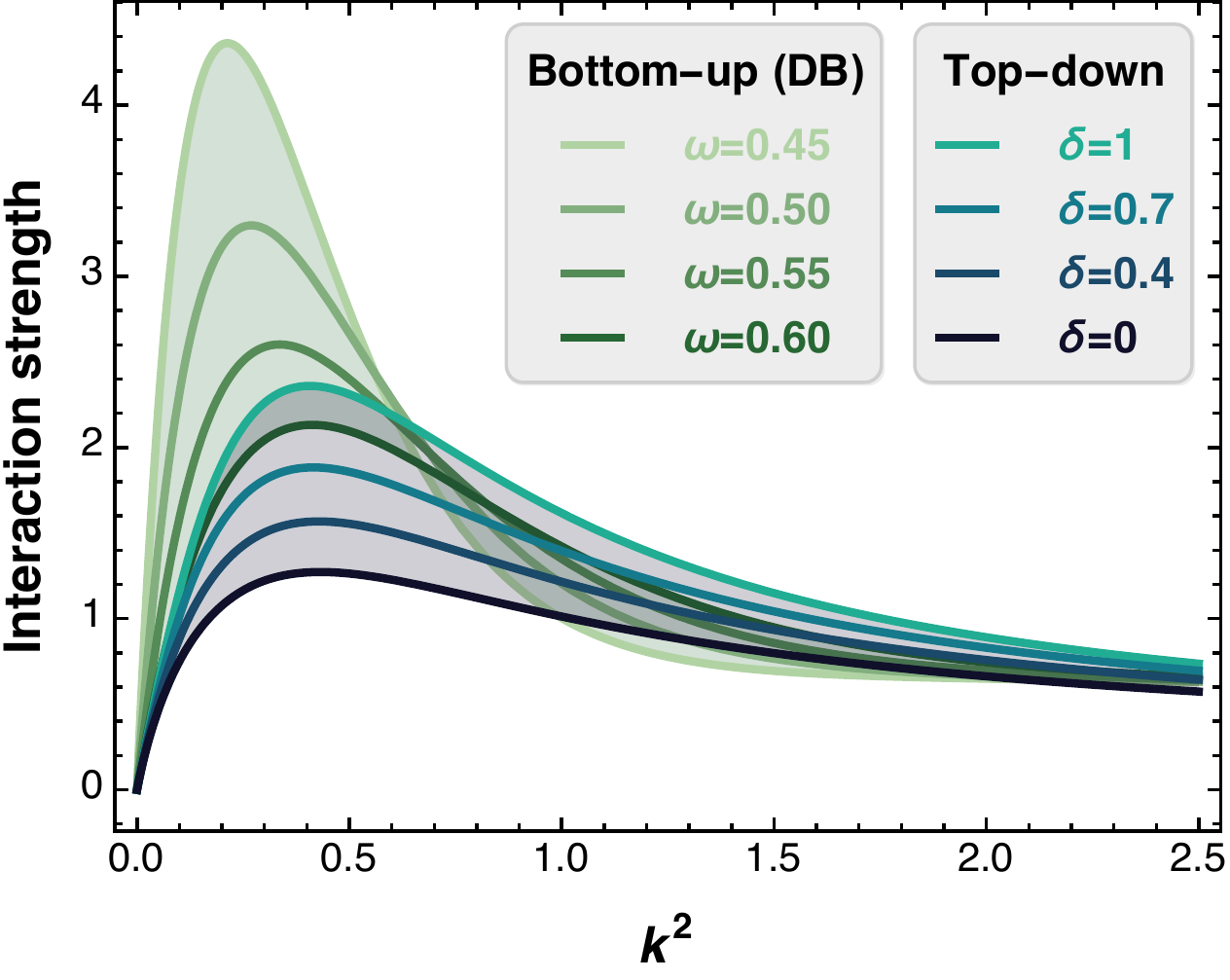}}
\caption{\label{fig:1}Comparison between top-down results for the gauge-sector interaction~[\1eq{td}, with those obtained using the bottom-up approach \1eq{bu}. \emph{Left panel}: RL truncation, ($\varsigma_\s{\rm RL}=0.87$ GeV) \emph{Right panel}: DB truncation, ($\varsigma_\s{\rm DB}=0.55$).  The bands denote either the domain of constant ground-state physics $\omega\in[0.45,0.6]$ GeV (bottom-up approach), or $\delta\in[0,1]$ (top-down approach). Notice in particular that one can see in this latter approach that the often employed approximation $L(k^2)\approx0$ is inaccurate, as $L$ shows a sizeable maximum in the intermediate momentum region, resulting in an enhancement of the interaction strength and an overall better agreement with the bottom-up interaction strength.}
\end{figure}

It is now possible to fully address the questions asked at the end of Sect.~\ref{sec-2}. Specifically, on the left (right) panel of Fig.~\ref{fig:1} we compare the top-down interaction strength ${\cal I}$ with the one obtained from a bottom-up approach using the RL (DB) truncation. 

It is immediately evident that while the RL interaction strength has the correct shape, it is way too large in the infrared. This is clearly an artefact of the truncation, as the bare vertex is too simple a description for capturing the complexity of hadronic phenomena: therefore, their satisfactory description can only be achieved in this case by artificially increasing the interaction strength. 

The bottom-up interaction strength obtained within the DB truncation scheme is instead in both qualitative as well as quantitative agreement with the top-down one. This represents quite a remarkable fact, meaning that the two approaches capture the same representation of the gauge and ghost sector dynamics, much beyond a generic acknowledgment of the dynamical emergence of non-zero and finite gluon mass-scale.  

\section{\label{sec-5}Conclusions}

Summarizing, our results shows that the interaction predicted by modern analyses of QCD's gauge sector is in near precise agreement with that required for a veracious description of measurable hadron properties using the most sophisticated matter-sector gap and Bethe-Salpeter kernels available today. This bridges a gap that had lain between nonperturbative continuum QCD studies and the {\it ab-initio} prediction of bound-state properties.
 
\begin{acknowledgement}
I thank the organizers of the ``21$^{\mathrm{st}}$ International Conference on Few-Body Problems in Physics'' for their kind invitation and hospitality, and all participants for providing a most stimulating and pleasant atmosphere.
\end{acknowledgement}



\end{document}